\documentclass[useAMS]{mn2e}
\usepackage{times}
\usepackage{rotating}
\long\def\symbolfootnote[#1]#2{\begingroup%
\def\thefootnote{\fnsymbol{footnote}}\footnote[#1]{#2}\endgroup}
\newcommand{\gae}{\lower 2pt \hbox{$\, \buildrel {\scriptstyle >}\over {\scriptstyle
\sim}\,$}}
\newcommand{\lae}{\lower 2pt \hbox{$\, \buildrel {\scriptstyle <}\over {\scriptstyle
\sim}\,$}}

\begin{document}

\title[GRB high energy photons detected by Fermi]
{External forward shock origin of high energy emission for three GRBs detected by \it{Fermi}}

\author[P. Kumar \& R. Barniol Duran]{P. Kumar$^{1}$\thanks
{E-mail: pk@astro.as.utexas.edu, rbarniol@physics.utexas.edu}
and R. Barniol Duran$^{1,2}$\footnotemark[1] \\
$^{1}$Department of Astronomy, University of Texas at Austin, Austin, TX 78712, USA\\
$^{2}$Department of Physics, University of Texas at Austin, Austin,
TX 78712, USA}

\date{Accepted 2010 June 30;
      Received 2010 June 29;
      in original form 2009 October 27}

\pubyear{2010}

\maketitle

\begin{abstract}

We analyze the $>$100 MeV data for 3 GRBs detected by the {\it Fermi} satellite (GRBs 080916C, 
090510, 090902B) and find that these photons were generated via synchrotron 
emission in the external forward shock. We arrive at this conclusion by 
four different methods as follows. (1) We check the light curve and spectral 
behavior of the $>$100 MeV data, and late time X-ray and optical data, 
and find them consistent with the so called closure relations for the 
external forward shock radiation. (2) We calculate the expected 
external forward shock synchrotron flux at 100 MeV,
which is essentially a function of the total energy in the burst alone, and 
it matches the observed flux value. (3) We determine the external forward shock 
model parameters using the $>$100 MeV data (a very large phase space of
parameters is allowed by the high energy data alone), and for each point in
the allowed parameter space we calculate the expected X-ray 
and optical fluxes at late times (hours to days after the burst) and find
these to be in good agreement with the observed data for the entire parameter
space allowed by the $>$100 MeV data. (4) We calculate the external forward 
shock model parameters using only the late time X-ray, optical and radio data 
and from these estimate the expected flux at $>$100 MeV at the end
of the sub-MeV burst (and at subsequent times) and find that to be
entirely consistent with the high energy data obtained by {\it Fermi}/LAT. 
The ability of a simple external forward shock, with two 
empirical parameters (total burst energy and energy in electrons) and two
free parameters (circum-stellar density and energy in magnetic fields), 
to fit the entire data from the end of the burst (1--50s) to about a week, 
covering more than eight-decades in photon frequency --- $>$10$^2$MeV, 
X-ray, optical and radio --- provides compelling confirmation of 
the external forward shock synchrotron origin
of the $>$100 MeV radiation from these {\it Fermi} GRBs.  Moreover, the parameters 
determined in points (3) and (4) show that the magnetic field required in 
these GRBs is consistent with shock-compressed magnetic field in the 
circum-stellar medium with pre-shocked values of a few tens of micro-Gauss.
\end{abstract}


\section{Introduction}

The {\it Fermi} Satellite has opened a new and sensitive window in the study 
of GRBs (gamma-ray bursts); for a general review of GRBs see 
Gehrels, Ramirez-Ruiz \& Fox (2009), M\'esz\'aros (2006), 
Piran (2004), Woosley \& Bloom (2006), Zhang (2007). 
So far, in about one year of operation, {\it Fermi} has detected 12 GRBs 
with photons with energies $>$100 MeV. The $>$10$^2$ MeV emission 
of most bursts detected by the LAT (Large Area Telescope: energy coverage
20 MeV to $>$300 GeV) instrument aboard the {\it Fermi} satellite 
shows two very interesting features (Omedei et al. 2009):  (1) The first 
$>$100 MeV photon arrives later than the first lower energy photon 
($\lae$1 MeV) detected by GBM (Gamma-ray Burst Monitor),  (2) The $>$100
 MeV emission lasts for much longer time compared to the burst duration 
in the sub-MeV band (the light curve in sub-MeV band declines very rapidly).

\begin{table*}
\begin{center}
\begin{tabular}{ccccccc}
\hline
& $\beta_{LAT}$ & $p$ & $z$ & $d_{L28}$ & $t_{GRB}[s]$ & $E_{\gamma,iso}[erg]$  \\
\hline \hline
GRB 080916C  & $1.20 \pm 0.03$ & $2.4 \pm 0.06$ & $4.3$ & $12.3$ & $60$ & $8.8\times 10^{54}$\\
GRB 090510 & $1.1 \pm 0.1$ & $2.2 \pm 0.2$ & $0.9$ & $1.8$ & $0.3$ & $1.08\times10^{53}$\\
GRB 090902B & $1.1 \pm 0.1$ & $2.2 \pm 0.2$ & $1.8$ & $4.3$ & $30$ & $3.63\times10^{54}$\\
\hline
\end{tabular}
\end{center}
\caption{{\small The main quantities used in our analysis for these 3 GRBs. 
$\beta_{LAT}$ is the spectral index for the $>100$MeV data, $p$ is the 
power-law index for the energy distribution of injected electrons i.e. $dn/d\gamma\propto
\gamma^{-p}$, $z$ is the redshift, $d_{L28}$ is the luminosity distance 
in units of $10^{28}$cm, $t_{GRB}$ is the approximate burst duration in the 
{\it Fermi}/GBM band and $E_{\gamma, iso}$ is the isotropic 
equivalent of energy observed in $\gamma$-rays in the 10keV-10GeV band for 
GRB080916C and GRB090902B, and in the 10keV-30GeV band for GRB090510. Data taken from Abdo et al. (2009a, 2009b,
2009c), De Pasquale et al. (2010).}}
\end{table*}

There are many possible $>$100 MeV photons generation mechanisms 
proposed in the context of GRBs; see Gupta \& Zhang (2007) and 
Fan \& Piran (2008) for a review.  
Shortly after the observations of GRB 080916C (Abdo et al. 2009a), 
we proposed a simple idea: the $>$100 MeV photons in GRB 080916C are 
produced via synchrotron emission in the external forward shock 
(Kumar \& Barniol Duran 2009). This proposal naturally explains 
the observed delay in the peak of the light curve 
for $>$100 MeV photons -- it corresponds to 
the deceleration time-scale of the relativistic ejecta -- and also 
the long lasting $>$100 MeV emission, which corresponds to the 
power-law decay nature of the external forward shock (ES) emission
(the ES model was first proposed by Rees \& M\'esz\'aros 1992, 
M\'esz\'aros \& Rees 1993, Paczy\'nski \& Rhoads 1993; for a comprehensive review of the ES model, 
see, e.g., Piran, 2004, and references therein).
Following our initial analysis on GRB 080916C, a number of groups have provided
evidence for the external forward shock origin of {\it Fermi}/LAT observations (Gao et al. 2009;
Ghirlanda, Ghisellini, Nava 2010; Ghisellini, Ghirlanda, Nava 2010; 
De Pasquale et al. 2010).

In this paper we analyze the $>$100 MeV emission of GRB 090510 
and GRB 090902B in detail, and discuss the main results of our 
prior calculation for GRB 080916C (Kumar \& Barniol Duran 2009), 
to show that the high energy radiation for all these three 
arose in the external forward shock via the synchrotron process. 
These three bursts - one short and two long GRBs - are selected 
in this work because the high energy data
for these bursts have been published by the {\it Fermi} team as well as the fact
that they have good afterglow follow up observations 
in the X-ray and optical bands (and also the radio band for GRB 090902B)
to allow for a thorough analysis of data covering more than a factor
10$^8$ in frequency and $>10^4$ in time to piece together the high energy 
photon generation mechanism, and cross check this in multiple different
ways. 

In the next section (\S2) we provide a simple analysis of the LAT
spectrum and light curve for these three bursts to show that the
data are consistent with the external forward shock model. This analysis 
consists of verifying whether the temporal decay index and the
spectral index satisfy the relation expected for the ES emission
(closure relation), and comparing the observed flux in the LAT band
with the prediction of the ES model (according to this model the
high energy flux is a function of blast wave energy, independent of 
the unknown circum-stellar medium density, and extremely weakly dependent 
on the energy fraction in magnetic fields).

We describe in \S3 how the $>$100 MeV data alone can 
be used to theoretically estimate the emission at late times 
($t\gae$ a few hours) in the X-ray and optical bands within the 
framework of the external forward shock model, and that for these 
three bursts the expected flux according to the ES model is in 
agreement with the observed data in these bands.

Moreover, if we determine the ES parameters ($\epsilon_e$, $\epsilon_B$,
$n$, and $E$)\footnote{$\epsilon_e$ and $\epsilon_B$ are the energy 
fraction in electrons and magnetic field for the shocked fluid, $n$ is the
number density of protons in the burst circum-stellar medium, and 
$E$ is the kinetic energy in the ES blast wave.} using only 
the late time X-ray and optical fluxes (and radio data), we can predict 
the flux at $>$100 MeV at any time after the deceleration time for the GRB
relativistic outflow. We show in \S3 that this predicted flux at 
$>10^2$MeV is consistent with the value observed by the {\it Fermi} satellite
for the bursts analyzed in this paper.

These exercises and results show that the high energy emission is due 
to the external shock as discussed in \S3. We also describe in \S3 
that the magnetic field in the shocked fluid --- responsible for the
generation of $>$ 100 MeV photons as well as the late time X-ray and
optical photons via the synchrotron mechanism --- is consistent with the
shock compression of a circum-stellar magnetic field of a few
tens of micro-Gauss.

It is important to point out that we do not consider in this work
the prompt sub-MeV emission mechanism for GRBs --- which is well known to have
a separate and distinct origin as evidenced by the very rapid decay of 
sub-MeV flux observed by Swift and {\it Fermi}/GBM (the flux in the sub-MeV
band drops-off with time as $\sim t^{-3}$ or faster as opposed to the 
$\sim t^{-1}$ observed in the LAT band). Nor do we investigate the
emission process for photons in the LAT band during the
prompt burst phase.

\section{ES model and the $>$100 MeV emission from GRBs: Simple arguments}

In this work we consider 3 GRBs detected by {\it Fermi}/LAT in the 
$>10^2$ MeV band: GRB 080916C (Abdo et al. 2009a), GRB 090510 (Abdo
et al. 2009b, De Pasquale et al. 2010) and GRB 090902B (Abdo et al. 2009c).  
These bursts show the ``generic'' features observed in the $>$100 MeV emission 
of most of {\it Fermi} GRBs mentioned above, and these are the only three bursts
for which we have optical, X-ray and {\it Fermi} data available. Some basic 
information for these 3 GRBs have been summarized in Table 1.    

The synchrotron process in the ES model predicts a relationship between 
the temporal decay index ($\alpha$) of the light curve and the energy 
spectral index ($\beta$), which are so called closure relations. These 
relations serve as a quick check for whether or not the observed 
radiation is being produced in the external shock. 
 In this paper, we use the convention 
$f(\nu,t) \propto \nu^{-\beta}t^{-\alpha}$.    

Since the {\it Fermi}/LAT band detects very high energy photons ($\gae10^2$MeV), 
it is reasonable to assume that this band lies above all the synchrotron 
characteristic frequencies (assuming that the emission process is
synchrotron). In this case the spectrum should be $\propto \nu^{-p/2}$  (Sari, Piran, Narayan 1998)---
where $p$ is the power law index of the injected electrons' energy distribution ---
and according to the external forward shock model (see, e.g., Panaitescu \& Kumar 2000), the light curve should decay as 
$\propto t^{-(3p-2)/4}$, giving the following closure relation: 
$\alpha=(3\beta-1)/2$. Using the data in Table 1 we find that all three
bursts satisfy this closure relation (Table 2), which encourages us to 
continue our diagnosis of the $>$100 MeV emission in the context of the 
ES model.

\begin{table*}
\begin{minipage}{\textwidth}
\begin{center}
\begin{tabular}{c|cc|ccc}
\hline
& $\alpha_{ES}$ & $\alpha_{obs}$ & $t[s]$ & $f_{100MeV}^{ES}$\symbolfootnote[1]{Fluxes in this table are in $n$Jy.  
The fluxes are calculated using equation (1), the data in Table 1, 
and setting the isotropic kinetic energy in the ES to be $E=E_{\gamma,iso}$, 
which gives a lower limit on $E$; most likely $E = {\rm few}\, \times E_{\gamma,iso}$ and we find that 
using $E \sim 3 \times E_{\gamma,iso}$ the fluxes match the observed values very well.
Also, for this calculation, $\epsilon_B = 10^{-5}$, $\epsilon_e = 0.25$, $p=2.4$ and $Y < 1$.} & $f_{100MeV}^{obs}$  \\
\hline \hline
GRB 080916C  &  $1.30 \pm 0.05$ & $1.2 \pm 0.2$ & $150$ & $>16$ & $67$\\
GRB 090510 &  $1.2 \pm 0.2$ & $1.38 \pm 0.07$ & $100$ & $>3$ & $14$\\
GRB 090902B &  $1.2 \pm 0.2$ & $\sim 1.5$ & $50$ & $>100$ & $220$\\
\hline
\end{tabular}
\end{center}
\end{minipage}
\caption{{\small Comparison between the temporal decay index ($\alpha_{ES}$) 
expected for the external forward shock model, and the observed decay index 
($\alpha_{obs}$); these values are equal to within 1--$\sigma$ error bar. 
The ES flux calculated at time $t$ is compared to the observed value at 
the same time. These two values are also consistent, further lending
support to the ES origin of the $>$ 100 MeV emission. Data are obtained 
from the same references as in Table 1. The theoretically calculated
flux would be larger if $\epsilon_e>0.25$; GRB afterglow data for 8 
well studied bursts suggest that $0.2<\epsilon_e\lae0.8$ (Panaitescu 
\& Kumar 2001).}}
\end{table*}

We next check to see if the predicted magnitude of the synchrotron 
flux in the ES is consistent with the observed values. This calculation 
would seem very uncertain at first, but we note that the predicted external 
forward shock synchrotron flux at a frequency larger than all characteristic 
frequencies of the synchrotron emission is independent of the circum-stellar
medium (CSM) density, $n$, and it is extremely weakly dependent on the 
fraction of the energy of the shocked gas in the magnetic field, $\epsilon_B$,
which is a highly uncertain parameter for the ES model.
The density falls off as $\propto R^{-s}$, where $R$ is the distance
from the center of the explosion, and $s=0$ corresponds to a constant CSM
and $s=2$ corresponds to a CSM carved out by the progenitor star's wind.
The flux is given by (see e.g. Kumar 2000, Panaitescu \& Kumar 2000):
\begin{eqnarray}
 f_{\nu} &=& {0.2 \rm mJy}\, E_{55}^{{p+2\over4}} \epsilon_e^{p-1} 
 \epsilon_{B,-2}^{{p-2\over4}} t_1^{-{3p-2\over4}} 
  \nu_8^{-{p\over 2}} (1+Y)^{-1} \nonumber \\
   && \times (1+z)^{{p+2\over4}} d_{L28}^{-2}, 
\end{eqnarray} where $\epsilon_e$ is the fraction of energy of the
shocked gas in electrons, $t_1=t/10s$
is the time since the beginning of the explosion in the observer frame (in
units of 10s), $\nu_8$ is photon energy in units of 100MeV, 
$E_{55}= E/10^{55}$erg is the scaled isotropic kinetic energy 
in the ES, $Y$ is the Compton-$Y$ parameter,
$z$ is the redshift and $d_{L28}$ is the luminosity distance to the burst 
(in units of $10^{28}$cm). Using the values of Table 1, we can predict 
the expected flux at 100 MeV from the ES and compare it to the observed value 
at the same time. We show in Table 2 that the observed high energy
flux is consistent with the theoretically expected values for all three
busts.

The fact that these bursts satisfy the closure relation, and that the observed
$>10^2$MeV flux is consistent with theoretical expectations, suggests that 
the high energy emission detected by {\it Fermi}/LAT from GRBs is produced via 
synchrotron emission in the ES. In the next section we carry out a more
detailed analysis that includes all the available data from these
bursts during the ``afterglow" phase, i.e. after the emission in the 
sub-MeV band has ended (or fallen below {\it Fermi}/GBM threshold).

\section{Detailed synthesis of all available data and the external forward shock model}

The simple arguments presented in the last section provide tantalizing
evidence that the high energy photons from the three bursts considered
in this paper are synchrotron photons produced in the external forward shock.
We present a more detailed analysis in this section where we consider
all available data for the three bursts after the end of the emission
in the {\it Fermi}/GBM band, i.e. for $t>t_{GRB}$, where $t_{GRB}$ is the ``burst
duration" provided in Table 1. The data we consider 
consist of $>10^2$MeV emission observed by {\it Fermi}/LAT and AGILE/GRID, 
X-ray data from Swift/XRT, optical data from Swift/UVOT and various
ground based observatories, and radio data from Westerbork in the case
of GRB 090902B.

The {\it main idea} is to use the $>10^2$MeV data to constrain the ES parameters
($\epsilon_e$, $\epsilon_B$, $n$ and $E$)\footnote{In addition to these
four parameters, the ES model also has an extra two, which are $s$ and $p$.  However, these 
last two can be estimated fairly directly by looking at the spectrum 
and temporal decay indeces of the light curves at different wavelengths.} --- which as we shall see 
allow for a large hyper-surface in this space --- and for each of the points
 in the allowed 4-D parameter space calculate the flux in the X-ray, optical 
and radio bands from the external forward shock at those times where data in 
one of these bands are available for comparison with the observed value. It would be 
tempting to think that such an exercise cannot be very illuminating as the ES
flux calculated at any given time in these bands would have a large
uncertainly that would reflect the large volume of the sub-space of 4-D 
parameter space allowed by the $>$10$^2$MeV data alone. This, however, turns
out to be incorrect -- the afterglow flux generated by the ES in the X-ray 
and optical bands (before the time of jet break) is almost uniquely 
determined from the high-energy photon flux; the entire sub-space of the
4-D space, allowed by the $>$10$^2$MeV data, projects to an extremely small 
region (almost to a point) as far as
the emission at any frequency larger than $\sim\nu_i$ is concerned;  
$\nu_i$ is the synchrotron frequency corresponding to the minimum energy of 
injected electrons (electrons just behind the shock front), which we also refer to as synchrotron injection frequency.
Therefore, we can compare the ES model predictions of flux in the X-ray and
optical bands with the observed data, and either rule out the
ES origin for high energy photons or confirm it\footnote{It should be 
pointed out that the X-ray afterglow light curves of long-GRBs are 
rather complicated during the first few hours (see e.g. Nousek et al. 2006,  
O'Brien et al. 2006) and the ES model in its simplest form 
can't explain these features, however the behavior becomes simpler and consistent 
with ES origin after about 1/2 day.}. 

We also carry out this exercise in the {\it reverse direction}, i.e. find the
sub-space of 4-D parameter space allowed by the late time ($t\gae$1day)
X-ray, optical, and radio data, and then calculate the expected $>$10$^2$MeV
flux at early times for this allowed subspace for comparison with the 
observed {\it Fermi}/LAT data. This {\it reverse direction} exercise is not 
equivalent to the one described in the preceding paragraph since the 
4-D sub-space allowed by the $>$10$^2$MeV data
and that by the late time X-ray and optical data are in general quite
different (of course they have common points whenever early high-energy
and late low energy emissions arise from the same ES).

The input physics in all of these calculations consist of the following main 
ingredients: synchrotron frequency and flux (see Rybicki \& Lightman 1979
for detailed formulae; a convenient summary of the relevant equations can also
be found in Kumar \& Narayan 2009), Blanford-McKee self-similar solution 
for the ES (Blandford \& McKee 1976), electron cooling due to synchrotron and synchrotron self-Compton
radiation (Klein-Nishina reduction to the cross-section is very 
important to incorporate for all the three bursts for at least a fraction
of the 4-D parameter space), and the emergent synchrotron spectrum as
described in e.g. Sari, Piran \& Narayan (1998). Although the calculations
we present in the following sections can be carried out analytically
(e.g. Kumar \& McMahon 2008), it is somewhat tedious, and so we have coded
all the relevant physics in a program and use that for finding the allowed
part of 4-D parameter space and for comparing the results of theoretical
calculation with the observed data. Numerical codes have also the 
advantage that they enable us to make fewer assumptions and approximations.
Nevertheless, we present a few analytical estimates 
to give the reader a flavor of the calculations involved.

We analyze the data for each of the three bursts individually in the 
following three sub-sections in reverse chronological order.

\subsection{GRB 090902B}

The {\it Fermi}/LAT and GBM observations of this burst can be found in 
Abdo et al. (2009c). The X-ray data for this GRB started at about 
half a day after the trigger time. The spectrum in the 0.3--10 keV X-ray 
band was found to be $\beta_x = 0.9 \pm 0.1$, and the light curve decayed as 
$\alpha_x=1.30\pm0.04$ (Pandey et al. 2010). The optical observations by 
Swift/UVOT started at the same time (Swenson \& Stratta 2009) and show 
$\alpha_{opt} \sim 1.2$.  ROTSE also 
detected the optical afterglow starting at $\sim 1.4$ hours and 
its decay is consistent with the UVOT decay (Pandey et al. 2009).
The Faulkes Telescope North also observed the afterglow at 
about $21$ hours after the burst using the R filter (Guidorzi et al. 2009).
There is a radio detection available at about 1.3 days after the
burst and its flux is $\sim111\mu$Jy at $4.8$GHz (van der Host et al. 2009).

The late time afterglow data obtained by Swift/XRT show that the X-ray
band, 0.3--10 keV ($\nu_x$), should lie between $\nu_i$ (the synchrotron 
injection frequency) and $\nu_c$ (the synchrotron 
frequency corresponding to the electrons' energy for which the radiative loss time-scale
equals the dynamical time; we also refer to it as synchrotron cooling frequency).
This is because $\nu_x>\nu_i$, otherwise the light curve would be
rising with time instead of the observed decline. 
Moreover, if $\nu_c < \nu_x$, then $p=2\beta_x\sim1.8\pm0.2$, and in
that case $f_{\nu_x}(t)\propto t^{-(3p+10)/16}=t^{-0.96\pm0.04}$, and that 
is inconsistent with the observed decline of the X-ray light curve 
(for decay indeces for values of $p<2$ see the table 1 in 
Zhang \& M\'esz\'aros (2004)). Thus, $\nu_i < \nu_x < \nu_c$, 
so that $\beta_x = (p-1)/2$ or $p\sim2.8\pm0.2$.

Next we determine if the X-ray data are consistent with a constant density 
circum-stellar medium or a wind-like medium. For $s=0$ 
($s=2$) the expected temporal decay index of the X-ray light curve is 
$\alpha=3(p-1)/4=1.35\pm0.15$ ($\alpha=(3p-1)/4=1.85\pm0.15$). 
Thus, a constant density 
circum-stellar medium is favored for this GRB.  

The XRT flux at 1keV at $t=12.5$hr was reported to be $0.4\mu$Jy 
(Pandey et al.  2010). Extrapolating this flux to the optical band
using the observed values of $\alpha_x$ and $\beta_x$ we find the
flux at 21hr to be within a factor $\sim 3$ of the $\sim15\mu$Jy flux reported by
Guidorzi et al. (2009). Thus, the emissions in the optical and the
X-ray bands arise in the same source (ES) with $\nu_i$ below the 
optical band; also, if the optical band were below $\nu_i$, then 
the optical light curve would be increasing with time, which is not 
observed. Moreover, the optical and X-ray data together 
provide a more accurate determination of the spectral index 
to be $0.69\pm0.06$ or $p=2.38\pm0.12$ which is consistent with
the $p$-value for the high-energy data at $t>t_{GRB}$ (see Table 1).

If the $>$10$^2$MeV emission is produced in the external forward shock
then we should be able to show that the early high energy $\gamma$-ray
flux is consistent with the late time X-ray and optical data.  
We first show this approximately using analytical calculations, and
then present results obtained by a more accurate numerical calculation
in our figures. 

The observed flux at 100 MeV and $t=50$s can be extrapolated to half a day to
estimate the flux at 1 keV. This requires the knowledge of where $\nu_c$
lies at this time.  It can be shown that $\nu_c\sim100$ MeV at 50 s
in order that the flux at 100 keV does not exceed the observed flux limit
(see subsection below).  Therefore, $\nu_c\propto t^{-1/2}$ is $\sim 3$ MeV
at 12.5 hr, and thus the expected flux at 1 keV is $\sim 0.5 \mu$Jy which agrees
with the observed value.  Therefore, we can conclude that the $>$100MeV, 
X-ray and optical photons were all produced by the same source, and we suggest 
that this source must be the external forward shock as already determined 
for the X-ray and optical data. 

We now turn to determining the ES parameter space for this burst.  We can 
determine this space using both the forward direction and reverse direction approaches.
We first list the constraints on the ES model, 
then give a few analytical estimates using the equations in, 
for example, Panaitescu \& Kumar (2002), 
and then present the results of our detail numerical calculations.  

\subsubsection{Forward direction}

In this subsection we only use the early high-energy emission to constrain
the ES parameter space.  The constraints at $t=50$ s are: 
(i) The flux at 100 MeV should agree with the observed value (see Table 2) - within
the error bar of 10\%, (ii) $\nu_c<100$MeV at 50s for consistency with the 
observed spectrum, (iii) the flux at 100 keV should be smaller than 0.04 mJy
(which is a factor of 10 less than the observed value), 
so that ES emission does not prevent the {\it Fermi}/GBM light curve to decay steeply 
after 25 seconds, and (iv) $Y<50$ so that the energy going into the second Inverse 
Compton is not excessive.  

The first 3 conditions give the following 3 equations at $t=50$ s.  
The cooling frequency is given by (Panaitescu \& Kumar 2002)

\begin{equation}
\nu_c \sim {6 \rm eV}\, E_{55}^{-1/2} n^{-1} \epsilon_{B,-2}^{-3/2} (1 + Y)^{-2} < {100 \rm MeV}\,.
\end{equation}
The flux at 100 keV, which is between $\nu_i$ and $\nu_c$ as discussed 
above, is (Panaitescu \& Kumar 2002)

\begin{equation}
f_{100 keV} \sim {53 \rm mJy}\, E_{55}^{1.35} n^{0.5} \epsilon_{B,-2}^{0.85} \epsilon_{e,-1}^{1.4} < {0.04 \rm  mJy}\,.
\end{equation}
And lastly, using (1), the flux at 100 MeV, which we assume is above $\nu_c$, is 
\begin{equation}
f_{100 MeV} \sim {1 \times 10^{-4} \rm mJy}\, E_{55}^{1.1} \epsilon_{B,-2}^{0.1} \epsilon_{e,-1}^{1.4} = {220 \rm  nJy}\,.
\end{equation}
Solving for $n$ from (3) and for $\epsilon_e$ from (4), and substituting in (2), we 
find that at 50 s $\nu_c \gae 50$ MeV. The injection frequency can also be estimated 
at $t=50$ s, it is given by 

\begin{equation}
\nu_i \sim {8 \rm keV}\, E_{55}^{1/2} \epsilon_{B,-2}^{1/2} \epsilon_{e,-1}^{2},
\end{equation}
and using (4), one finds 
$\nu_i \sim {25 \rm keV}\, E_{55}^{-1.07} \epsilon_{B,-2}^{0.36}$, which 
gives $\nu_i \sim 2$ keV for $\epsilon_{B}\sim10^{-5}$.  
These values of $\nu_i$ and $\nu_c$ are consistent with the 
values obtained with detail numerical calculations and reported 
in the Fig. 1 caption. 

Using (2), we can find a lower limit on $\epsilon_B$, which is given by
\begin{equation}
\epsilon_B \gae \frac{1\times10^{-7}}{n^{2/3} E_{55}^{1/3} (1+Y)^{4/3}}.
\end{equation} 
Also, we can solve for $\epsilon_e$ using (4) and substitute that into (3) 
to obtain an upper limit on $\epsilon_B$, which is
\begin{equation}
\epsilon_B \lae \frac{3\times10^{-7}}{n^{2/3} E_{55}^{1/3} (1+Y)^{4/3}}.
\end{equation}
Note that these estimates are consistent with the numerical results we present 
in Fig. 1.  We also find the $\epsilon_B \propto n^{-2/3}$ dependence 
that is shown in the figure. 

Moreover, with these parameters we can predict the fluxes at late times.
The X-ray and optical band lie between $\nu_i$ and 
$\nu_c$  at $\sim$ 1 day (see above).  The first X-ray data point is at 12.5 hr, and 
the theoretically expected flux at 1 keV at this time is given by

\begin{equation}
f_{1 keV} \sim {1 \rm mJy}\, E_{55}^{1.35} n^{0.5} \epsilon_{B,-2}^{0.85} \epsilon_{e,-1}^{1.4},
\end{equation}
and the optical flux at $\sim 7.5 \times 10^4$ s is 

\begin{equation}
f_{2 eV} \sim {47 \rm mJy}\, E_{55}^{1.35} n^{0.5} \epsilon_{B,-2}^{0.85} \epsilon_{e,-1}^{1.4}.
\end{equation}  
We can use (3) to find an upper limit for the X-ray and optical fluxes.  
In addition, we can find $\epsilon_e$ using (4), and use (6) to find a lower 
limit for these fluxes. We find that 

\begin{equation}
{0.5 \rm \mu Jy}\, \lae f_{1 keV} \lae {0.8 \rm \mu Jy}\,
\end{equation}
for the X-ray flux at 12.5 hr, and 

\begin{equation}
{25 \rm \mu Jy}\, \lae f_{2 eV} \lae {36 \rm \mu Jy}\,
\end{equation} for the optical flux at $\sim 7.5 \times 10^4$ s.  These 
estimates agree very well with the observed values of $0.4 \mu$ Jy
(Pandey et al. 2010) and $15 \mu$Jy (Guidorzi et al. 2009) at the 
respective bands and times.  We note that, although Inverse Compton cooling
is very important at late times, the X-ray band lies below $\nu_c$ and therefore
X-ray and optical fluxes are unaffected by Inverse Compton cooling. 

\begin{figure}
\begin{centering}
\centerline{\hbox{\includegraphics[width= 6cm,angle = 0]{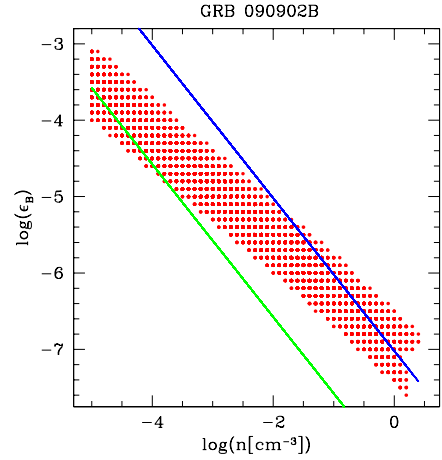}}}
\end{centering}
\caption{We determine the sub-space of the 4-D parameter space (for the 
forward external forward shock model with $s=0$) allowed by the high energy 
data for GRB 090902B at t=50s as described in \S3.1. The projection of the 
allowed subspace onto the $\epsilon_B$--$n$ plane is shown in this figure 
(dots); the discrete points reflect the numerical resolution of our 
calculation.  We also plot the expected $\epsilon_B$ for a shock compressed 
CSM magnetic field of 5 and 30 $\mu$-Gauss as the green and blue lines
respectively; for a CSM field of strength $B_0$, the value of $\epsilon_B$
downstream of the shock-front resulting from the shock compressed CSM
field is $\approx B_0^2/(2\pi n m_p c^2)$, where $n m_p$ is the CSM mass
density, and $c$ is the speed of light. Note that
no magnetic field amplification is needed, other than shock compression 
of a CSM magnetic field of $\sim30\mu$G, to produce the $>$100 MeV photons. 
The synchrotron injection and cooling frequencies at $t=50$s for the
sub-space of 4-D parameter space allowed by the high energy data are 
100eV$\lae\nu_i\lae$3keV and 30MeV$\lae\nu_c\lae$100 MeV respectively,
the Lorentz factor of the blast wave at $t=50$s lies between 330 and 1500, 
and $10^{55}$erg$\lae E \lae 3\times10^{55}$erg.
Note that at 0.5 day $\nu_i$ would be below the optical band, and $\nu_c>1$MeV,
and these values are consistent with the X-ray spectrum and the X-ray and 
optical decay indices at this time. }
\label{090902B-epsilonB} 
\end{figure}

\begin{figure*}
\begin{centering}
\centerline{\hbox{\includegraphics[width=12cm, angle = 0, clip=true, viewport=.0in .0in 8in 4.5in]{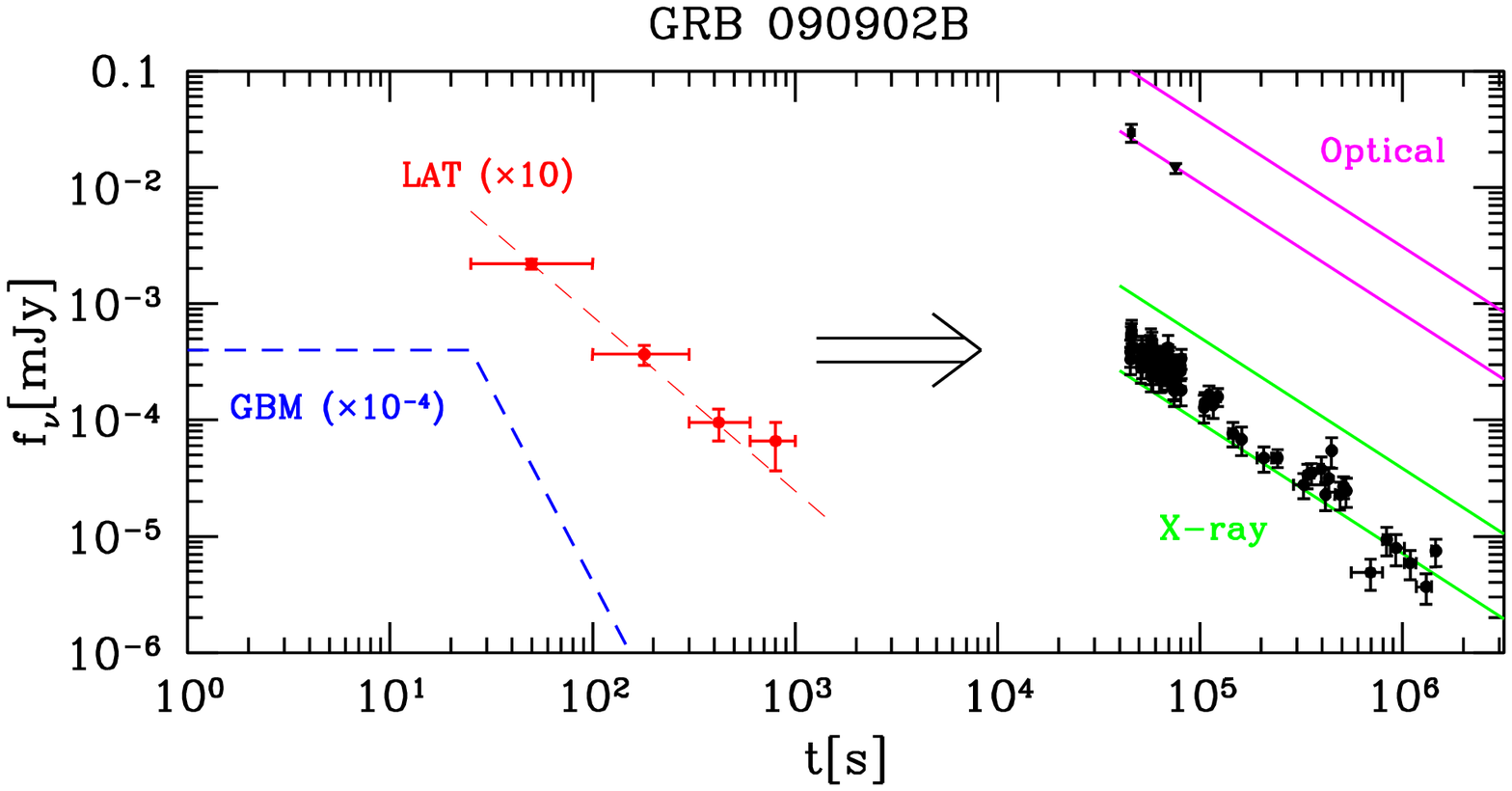}}}
\end{centering}
\caption { The optical and X-ray fluxes of GRB 090902B predicted at late times
using only the high energy data at 50s (assuming synchrotron emission
from external forward shock) are shown in the right half of this figure,
and the predicted flux values are compared with the observed data
(discrete points with error bars).
The width of the region between the green (magenta) lines indicates the uncertainty 
in the theoretically predicted X-ray (optical)
fluxes (the width is set by the error in the measurement of 100 MeV flux at
50s, and the error in the calculation of external forward shock flux due
to approximations made --
both these contribute roughly equally to the uncertainty in the predicted
flux at late times).
LAT (X-ray) data red (black) circles, are from Abdo et al. 2009c
(Evans et al. 2007, 2009) and were converted to flux density at 100 MeV (1keV)
using the average spectral index provided in the text.  Optical fluxes are from
Swenson et al. 2009 (square) and Guidorzi et al. 2009 (triangle) and were converted to 
flux density using $16.4$mag$\approx 1$mJy. The blue dashed
line shows schematically the light curve observed by {\it Fermi}/GBM. The predicted
value for the radio flux at one day has a very large range (not shown),
but consistent with the observed value. }  
\label{090902B-fwd}
\end{figure*}

\begin{figure}
\begin{centering}
\centerline{\hbox{\includegraphics[width=6cm,angle = 0]{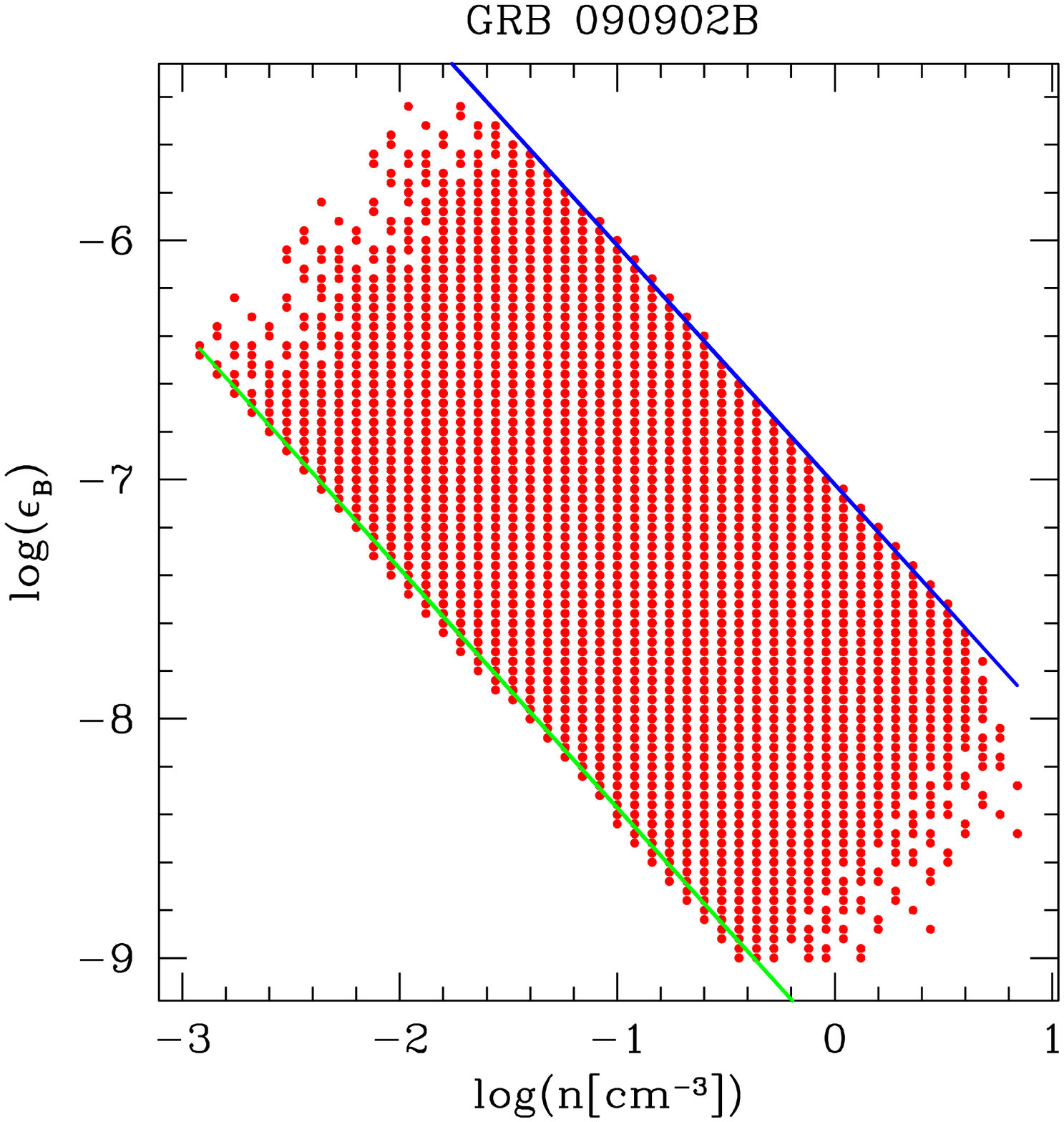}}}
\end{centering}
\caption {We determine the sub-space of the 4-D parameter space (for the 
external forward shock model with $s=0$) allowed by the late time ($t>0.5$day) 
X-ray, optical and radio data for GRB 090902B as described in \S3.1. 
The projection of the allowed subspace onto the $\epsilon_B$--$n$ plane 
at $t=0.5$day is shown in this figure (dots). We also plot the expected 
$\epsilon_B$ for a shock compressed CSM magnetic field of 
2 and 30 $\mu$-Gauss as the green and blue lines, respectively; for a 
CSM field of strength $B_0$, the value of $\epsilon_B$ downstream of the 
shock-front resulting from the shock compressed CSM field is 
$\approx B_0^2/(2\pi n m_p c^2)$, where $n m_p$ is the CSM mass
density, and $c$ is the speed of light. Note that 
no magnetic field amplification is needed, other than shock compression     
of a CSM magnetic field of strength $\lae30\mu$G, to produce the 
late time X-ray, optical and radio data. We arrived at this same conclusion
from the modeling of early time $>$100 MeV radiation alone (see fig. 
\ref{090902B-epsilonB}). }
\label{090902B-epsilonBlate}
\end{figure}


Next, we present the results obtained by detailed numerical 
calculations.  We use the same constraints described at the beginning of 
this subsection to determine the parameter space allowed by the high-energy 
early data. It is worth noting that in our numerical calculations throughout 
the paper we make no assumption regarding the ordering of the characteristic frequencies, 
nor the location of the observed bands with respect to them.
The projection of the sub-space of the 4-D parameter space allowed by the
high energy data onto the $\epsilon_B$--$n$ plane is shown in
Figure \ref{090902B-epsilonB}, and some of the other ES parameters are presented in 
the Fig. 1 caption. It is clear that there is a large sub-space
that is consistent with the LAT data, and also that the magnetic field
needed for the synchrotron source is consistent with the shock-compressed 
magnetic field in the CSM of strength $\lae30\mu$G. For each point 
in the 4-D space
allowed by the $>$10$^2$MeV data we calculate the X-ray and the optical flux at
late times. In spite of the fact that the 4-D sub-space allowed by the
LAT data is very large (fig. \ref{090902B-epsilonB}) the expected X-ray and 
optical flux at late times lie in a narrow range as shown by two 
diagonal bands in Figure \ref{090902B-fwd}; the width of these
bands has been artificially increased by a factor 2 to reflect the 
approximate treatment of the radial structure of the blast wave
and also to include in the calculation the effect of the  
blast wave spherical curvature on the ES emission (see, e.g., 
Appendix A of Panaitescu \& Kumar 2000; both of these effects together 
contribute roughly a factor of 2). We see that the observed X-ray and optical light curves lie within 
the theoretically calculated bands (Fig. \ref{090902B-fwd}).
This result strongly supports the ES model for the origin of the $>$10$^2$MeV 
photons.  

We note that the above mentioned extrapolation from early time, high-energy,
 data to late time, low-energy, flux prediction was carried out 
for a CSM with $s=0$. 
We have also carried out the same calculation but by assuming a wind medium ($s=2$), and
in this case we find that the expected flux at late times is smaller than the
observed values by a factor of $20$ or more; this conclusion is drawn 
by comparing the late optical and X-ray fluxes predicted at a single time with 
the observations at that same time, i.e. without making use of the temporal decay indices 
observed in these bands.  We pointed out above that the late time afterglow 
data for this burst are consistent with a uniform density medium, but not 
with a $s=2$ medium. Thus, 
there is a good agreement between the late time afterglow data and the 
early $>$10$^2$MeV data in regards to the property of the CSM; the two 
methods explore the CSM density at different radii.

\subsubsection{Reverse direction}

We carry out the above mentioned exercise in the reverse direction as 
well, i.e. we determine  the ES parameter space using only the late time 
X-ray, optical and radio data, and use these parameters to determine the flux 
at 10$^2$MeV at early times when {\it Fermi}/LAT observations were made. The constraints 
on ES model parameters at late times are the following:
(i) The X-ray and optical flux at 12.5 hr and $7.5 \times 10^4$ s, respectively, 
should match the ES flux at these bands and at these times, (ii) the radio 
flux at 1.3 d should be consistent with the observed value.  We first 
show some analytical estimates and then turn to more detailed numerical calculations.

Constraint (i) is simply equation (8) set equal to the observed value of $0.4 \mu$ Jy at 12.5 hr.
For the analytical estimates presented here, it is not necessary to use the optical flux at late times, since 
both the optical and X-ray bands lie between $\nu_i$ and $\nu_c$, so they 
provide identical constraints.  Constraint (ii), assuming that the radio frequency 
is below $\nu_i$, gives

\begin{equation}
f_{4.8 GHz} \sim {19 \rm mJy}\, E_{55}^{5/6} n^{1/2} \epsilon_{B,-2}^{1/3} \epsilon_{e,-1}^{-2/3} = {111 \rm \mu Jy}\,.
\end{equation}

Solving for $\epsilon_e$ in the last equation and substituting in constraint (i) 
gives an equation for $\epsilon_B$, which is

\begin{equation}
\epsilon_B = \frac{6\times10^{-8}}{n E_{55}^{2}}.
\end{equation} 
This estimate is consistent with the numerical result presented on Fig. \ref{090902B-epsilonBlate}.  
Moreover, one can see that we find $\epsilon_B \propto n^{-1}$, which is exactly 
what is found numerically (and agrees very well with the 
shock-compressed CSM field prediction).

We can now predict the high-energy flux at 100 MeV and early 
time using the ES parameters determined using late time afterglow data in 
X-ray and radio bands. We use (1) at $t=50$ s, substituting 
$\epsilon_e$ from (12) and $n$ from (13), and find that the flux should be

\begin{equation}
f_{100 MeV} \sim {200 \rm nJy}\, E_{55}^{3/4} \epsilon_{B,-5}^{-1/4},
\end{equation} 
in agreement with the observed value at $t=50$ s. We now turn to 
our numerical results.

\begin{figure*}
\begin{centering}
\centerline{\hbox{\includegraphics[width=12cm, angle = 0, clip=true, viewport=.0in .0in 8in 4.5in]{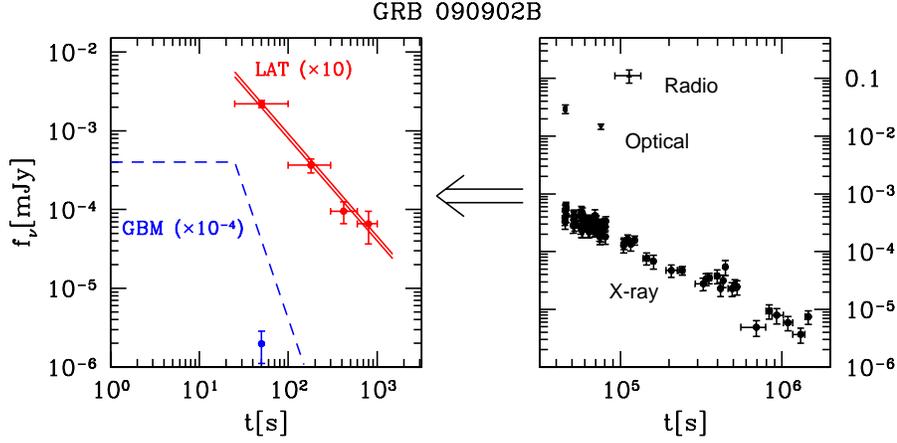}}}
\end{centering}
\caption {Using the X-ray, optical and radio data of GRB 090902B at late 
times (right panel) we constrain the external  forward shock parameters,
and then use these parameters to predict the 100 MeV flux at early times 
(left panel). The region between the red lines shows the range for the 
predicted flux at 100 MeV; note the remarkably narrow range for
the predicted 100 MeV flux in spite of the large spread to the allowed 
ES parameters as shown in fig. \ref{090902B-epsilonBlate}. 
 The blue point (left panel) indicates the flux at 100keV and 50s that 
we expect from the ES model; note that the ES flux at 100 keV falls well below 
the observed {\it Fermi}/GBM flux shown schematically by the dashed line in the 
left panel, and that is why the GBM light curve undergoes a rapid decline 
with time ($\sim t^{-3}$) at the end of the prompt burst phase.
The radio flux is taken from van der Host et al. 2009.  All other 
data are the same as in Figure \ref{090902B-fwd}.}
\label{090902B-reverse}
\end{figure*}

Using the same set of constraints presented at the beginning of this subsection, 
we perform our numerical calculations to determine the ES 
parameter space allowed by the late time ($t \gae 0.5$ d)
X-ray, optical and radio data and use that information to 
``predict'' the 100 MeV flux at early times ($t \lae 10^3$ s).
The numerical results of this exercise, for a $s=0$ CSM medium,
are also in good agreement with the {\it Fermi}/LAT data as shown in 
Figure \ref{090902B-reverse}.
Moreover, the flux from the ES at $t=50$s and 100 keV is found
to be much smaller than the flux observed by {\it Fermi}/GBM (Fig. 
\ref{090902B-reverse} - left panel), which is very reassuring, because otherwise this
would be in serious conflict with the steep decline of the light curve
observed in the sub-MeV band; this also shows that the sub-MeV and GeV 
radiations are produced by two different sources. 

We note that the range of values for $\epsilon_B$ allowed by the late time 
radio, optical \& X-ray afterglow data is entirely consistent with shock 
compressed circum-stellar medium magnetic field of strength $<30\mu$G 
(see fig. \ref{090902B-epsilonBlate}). We also point out that the afterglow
flux depends on the magnetic field $B$, and $B^2 \propto \epsilon_B n$, 
therefore, there is a degeneracy between $\epsilon_B$ and $n$ and that makes 
it very difficult to determine $n$ uniquely.

\subsection{GRB 090510}

The {\it Fermi}/LAT and GBM observations of this burst are 
described in Abdo et al. (2009b) and De Pasquale et al. (2010).
This short burst has very early X-ray and optical data
starting only 100s after the burst.  The X-ray spectrum 
is $\beta_x=0.57\pm0.08$ (Grupe \& Hoverstein 2009). The temporal decay 
index is $\alpha_{x,1}=0.74\pm0.03$ during the initial $\sim 10^3$s and 
subsequently the decay steepens to $\alpha_{x,2}=2.18\pm0.10$
with a break at $t_x=1.43^{+0.09}_{-0.15}$ks. The optical 
data shows $\alpha_{opt,1}=-0.5^{+0.11}_{-0.13}$ and 
$\alpha_{opt,2}=1.13^{+0.11}_{-0.10}$ with a break at 
$t_{opt}=1.58^{+0.46}_{-0.37}$ks (De Pasquale et al. 2010).

In the context of the ES model (also considered by Gao et al. 2009, 
Ghirlanda, Ghisellini, Nava 2010 and De Pasquale et al. 2010 for the 
case of this specific burst), the data suggests that 
$\nu_x < \nu_c$, because in this case $\beta_x = (p-1)/2$, so 
$p=2.14\pm0.16$ and the temporal decay index (for $s=0$) is 
$\alpha_x=3(p-1)/4=0.86\pm0.12$ consistent with the 
observed value of $\alpha_{x,1}$. If we take $\nu_x > \nu_c$, 
then $\beta_x=p/2$, so $p=1.14\pm0.16$ and the temporal decay 
index should have been $\alpha_x = (3p+10)/16=0.84\pm0.03$, 
since $p<2$, which is consistent with the observed temporal decay, 
however, the expected optical light curve index for this value of $p$ 
is $\alpha_{opt}=-(p+2)/(8p-8)=-2.8$, which is inconsistent 
with the observed value of $\alpha_{opt,1}$ (see next paragraph).
The X-ray afterglow data also shows that 
the medium in the vicinity of the burst must have been of 
constant density. This is because, for an $s=2$ medium, the expected
temporal decay of the X-ray flux, when $\nu_x < \nu_c$, is 
$\propto t^{-(3p-1)/4}=t^{-1.36}$ -- much steeper 
than the observed decline of $t^{-0.74}$ -- while for 
$s=0$ the expected decline is consistent with observations 
(Gao et al. 2009).  

\begin{figure}
\begin{centering}
\centerline{\hbox{\includegraphics[width= 6cm,angle = 0]{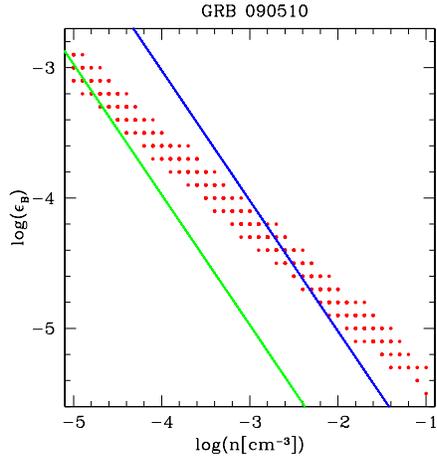}}}
\end{centering}
\caption {Using the observational constraints mentioned in the text (\S3.2), 
we determine the sub-space of 4-D parameter space (for the external forward shock
with $s=0$) allowed by the data
for GRB 090510 at $t=50$s. We show the projection of the allowed subspace onto the
$\epsilon_B$--$n$ plane in this figure (dots); the region agrees with the 
expected $\epsilon_B$ from shock-compressed CSM magnetic field of 
$\lae 30 \mu$G (the green and blue lines show $10\mu$G and $30\mu$G, 
respectively). The other parameters for the ES solution at this time
are: The Lorentz factor of the blast wave is between 260 and 970, 
$0.1<\epsilon_e<0.7$ and $10^{53}$erg$\lae E\lae 4\times10^{53}$ erg. 
At $t=100$ s, we also find $Y<4$, $\nu_i \sim 500$ eV, $\nu_c\sim 40$ MeV.} 
\label{090510-epsilonB}
\end{figure}

Given the fact that the break in the optical light curve and that in the X-ray light curve occur
at the same time, i.e. $t_x = t_{opt}$, it is unlikely that the emission 
in these two bands comes from two different, unrelated sources. Thus,
 it is natural to attribute both the optical and X-ray emissions to
the external forward shock. The fact that the optical light curve is
rising during the first $\sim0.5$hr as $t^{1/2}$ means that $\nu_{opt}<\nu_i$ during
this time period (Panaitescu \& Kumar 2000), where $\nu_{opt}$ is the optical band. 
The break seen in both light curves can be 
attributed to the jet break. The X-ray light curve decay of 
$t^{-2.2}$ for $t>1.4\times10^{3}$s agrees very well with the 
expected post-jet-break light curve of $\propto t^{-p} = t^{-2.12\pm0.14}$ 
(Rhoads 1999), and suggests a jet
opening angle of $\sim 1^o$ (Sari, Piran \& Halpern, 1999).
The reason that $\alpha_{opt,2}$ is not as steep 
can be understood the following way. At the time of the jet break, 
the optical band is below $\nu_i$, therefore, the light curve decays 
as $\propto t^{-1/3}$ instead of $\propto t^{-p}$ (Rhoads 1999). 
At later times, 
when $\nu_i$, which is decreasing rapidly, crosses the optical band, 
the optical light curve will transition slowly from
$\propto t^{-1/3}$ to $\propto t^{-p}$, and that is why $\alpha_{opt,2}$
is not as large as $\alpha_{x,2}$; the timescale for this transition
can be long/short depending on how far above $\gamma_i$ the asymptotic 
distribution of $n(\gamma)\propto \gamma^{-p}$ is attained.
This interpretation is supported by the results of our numerical 
calculation shown in Figure \ref{090510-epsilonB} --- we obtain a value 
of $\nu_i\sim500$eV at 100s, which should cross the optical band 
at $\sim 4000$s - a factor of $\sim3$ larger than the observed time of the 
jet break. This idea can 
be tested with optical data available at much later times: it should show 
the light curve slowly steepening to the asymptotic value of 
$\propto t^{-p}$. Moreover, the optical spectrum before 
the break in the light curve ($t<t_{opt}$) should be consistent with 
$\nu^{1/3}$. 

Is it possible that the rise of the optical band light curve might be due 
to the onset of the ES, while the initial X-ray emission
(until the break at $\sim 1.4$ ks) and the gamma-ray photons are 
from the ``internal shock'' mechanism (De Pasquale et al. 2010)? This seems 
unlikely, given that the density of the CSM required for the deceleration 
time of the GRB jet to be $\sim10^3$s ($t_{opt}$) is  
extremely low, as can be seen from the following equation

\begin{equation}
n = \frac{3E(1+z)^3}{32\pi c^5 m_p \Gamma^8 t_{peak}^3},
\end{equation} 
where $m_p$ is the mass of the proton, $c$ is the speed of light, 
$\Gamma$ is the initial Lorentz factor of the GRB jet, $t_{peak}$ is the time 
when the peak of the light curve is observed and $E$ is the isotropic 
energy in the ES. For GRB 090510, $\Gamma$ was determined to be $\Gamma \gae 10^3$
by using $\gamma \gamma$ opacity arguments (Abdo et al. 2009b), which is a 
limit applicable to the scenario proposed by De Pasquale et al. (2010),
where MeV and GeV photons are produced in the same source.  We take  
$t_{peak} \sim 10^3 s$ and $E \sim E_{\gamma,iso}$ and find that 
we need a CSM density of 
$n \approx 10^{-9} E_{53} \Gamma_{3}^{-8} t_{peak,3}^{-3} cm^{-3}$, which 
is much smaller than the mean density of the Universe at this redshift, and
therefore unphysical. Even though there is a strong dependence 
of CSM density on $\Gamma$, the upper limit on density provided above 
cannot be increased by more than a factor of $\sim 10$, since the error 
in the determination of $\Gamma$ is much less than a factor of 2 (Abdo
et al. 2009b). 
Thus, the possibility that the peak of the optical light curve
at $\sim10^3$s is due to the deceleration of the GRB jet seems very unlikely.
We note that in the scenario we present in this paper, the $>100$ MeV emission 
observed by {\it Fermi}/LAT and the lower 
energy ($\lae 1$ MeV) observed by {\it Fermi}/GBM are produced by two different 
sources, therefore, the pair-production argument can't be used to constrain 
$\Gamma$. However, in this scenario, the deceleration time for the GRB 
jet is $\lae 1$ s, and that means that the peak of the optical
light curve at $\sim 10^3$ s cannot correspond to the deceleration time.

We conclude that the available data suggest that optical and X-ray photons are 
coming from the same source (ES model). We now consider whether the 
observed $>$100 MeV emission is also consistent with the ES model.  We
first use the observed data to show that $>$100 MeV, X-ray and optical 
data are produced by the ES, then we provide some analytical estimates 
of the ES model parameters and later show the results of our 
detailed numerical results in the figures. 

AGILE/GRID reported a photon count in the 30 MeV--30 GeV band of 
$1.5\times10^{-3}$cm$^{-2}$ s$^{-1}$ at 10s, and the light curve was reported 
to decline as $t^{-1.3\pm0.15}$ (Giuliani et al. 2010). Therefore, the photon
flux at 100s in this band is estimated to be 
$\sim7.5\times10^{-5}$ cm$^{-2}$ s$^{-1}$ (Ghirlanda, Ghisellini, Nava, 2010; 
have also reported a single power-law decline of 
flux in the {\it Fermi} LAT band from $\sim$1s to 200s). The Swift/XRT reported 
a photon flux of 0.07 cm$^{-2}$ s$^{-1}$ in the 0.3--10 keV band at 100s. 
Using the spectrum reported in the Swift/XRT band (Grupe \& Hoverstein 2009) --- which 
is entirely consistent with the spectrum found in the AGILE/GRID
 band (Giuliani et al. 2010) --- to extrapolate the observed photon 
count in the XRT band to the GRID band we find the expected photon flux 
at 100s in the 30 MeV--30 GeV band of $7.9\times10^{-5}$ cm$^{-2}$ s$^{-1}$, 
and that is consistent with the flux observed by AGILE.

The peak of the optical light curve was observed at $\sim1000$ s
with a value of $\sim 100\mu$Jy, and the X-ray flux at 1000 s and $\sim4$keV was 
$2.2\mu$Jy (De Pasquale et al. 2010).
Since we attribute the optical light curve peak with the crossing of $\nu_i$
through the optical band, then the peak of the optical light curve 
determines the synchrotron flux at the peak of the spectrum.  Therefore, using 
the X-ray flux at 1000 s and the X-ray spectrum we can extrapolate
back to optical band (2 eV) and we find a flux of $170\mu$Jy, which is 
consistent, within a factor of better than 2, with the observed optical value 
at this time.  Therefore, we can conclude that $>$100 MeV, X-ray and optical emissions
are all produced by the same source, and that source must be the 
external forward shock as that is known to produce long lasting 
radiation in the X-ray and optical bands with a well known closure relation
between $\alpha$ and $\beta$ that is observed in GRB 090510 in all
energy bands.

Using the data in the LAT, XRT and optical bands we can determine the 
ES parameters for GRB 090510. The following observational constraints must 
be satisfied by the allowed ES parameters: (i) The flux at 100 MeV and 100s 
should be equal to the observed value (Table 2), (ii) $\nu_c<100$ MeV at 100s, 
(iii) the X-ray flux at 1000s and $\sim4$keV should be equal to the observed 
value of $2.2\mu$Jy (De Pasquale et al. 2010), and (iv) the flux at the peak 
of synchrotron spectrum should be $\sim 100\mu$Jy (De Pasquale et al. 2010). 
This last constraint arises because the optical flux peaks when $\nu_i$ 
passes the optical band, and therefore the peak synchrotron flux should be 
equal to the measured peak optical flux; it should be noted that the peak 
synchrotron flux for $s=0$ according to the ES model does not change
with time as long as the shock front moves at a relativistic speed.

We present some analytical estimates for the ES parameters before 
showing our detailed numerical results.  The ES flux at 100 MeV and $t=100$ s, assuming 
that 100 MeV is above $\nu_c$ is given by (1) and is

\begin{equation}
f_{100 MeV} \sim {2.4 \times 10^{-6} \rm mJy}\, E_{53}^{1.1} \epsilon_{B,-2}^{0.1} \epsilon_{e,-1}^{1.4} = {14 \rm  nJy}\,,
\end{equation}
which is the constraint (i).  The flux at 4 keV and 1000s, assuming that it is 
between $\nu_i$ and $\nu_c$ is given by 

\begin{equation}
f_{4 keV} \sim {3 \rm mJy}\, E_{53}^{1.35} n^{0.5} \epsilon_{B,-2}^{0.85} \epsilon_{e,-1}^{1.4} = {2.2 \rm \mu Jy}\,,
\end{equation}
which is constraint (iii).  And lastly, constraint (iv) is that the peak 
synchrotron flux should equal the flux at the peak of the optical 
light curve, i.e., 

\begin{equation}
f_{p} \sim {12 \rm mJy}\, E_{53} n^{1/2} \epsilon_{B,-2}^{1/2} = {100 \rm \mu Jy}\,.
\end{equation}

Just as was done for GRB090902B, constraint (ii) gives a lower 
limit on $\epsilon_B$, which in the case for this GRB is not too useful.
Instead, we can solve $\epsilon_e$ from (16) and substitute it in (17), 
which gives

\begin{equation}
\epsilon_B = \frac{1\times10^{-6}}{E_{53}^{1/3} n^{2/3} (1+Y)^{4/3}},
\end{equation} consistent with the numerical calculation presented 
in Fig. 5. Also, with this last expression and using (18) we find 
that the CSM density for this GRB is 

\begin{equation}
n \sim {0.3 \rm cm^{-3}}\, (1+Y)^4 E_{53}^{-5},
\end{equation} which is also consistent with the fact that we only 
find numerical solutions with CSM densities lower than 
$\sim$0.1 cm$^{-3}$.  

For the ES parameters of this burst, the cooling frequency at 100 s 
can be estimated to be 

\begin{equation}
\nu_c \sim {76 \rm eV}\, E_{53}^{-1/2} n^{-1} \epsilon_{B,-2}^{-3/2} (1 + Y)^{-2},
\end{equation} and substituting $n$ from (18) gives 
$\nu_c \sim {1 \rm MeV}\, E_{53}^{3/2} \epsilon_{B,-2}^{-1/2} (1 + Y)^{-2}$.
Thus, for $\epsilon_B \sim 10^{-5}$ we find 
$\nu_c \sim 30$ MeV.  The injection frequency at 100 s is 
given by 

\begin{equation}
\nu_i \sim {240 \rm eV}\, E_{53}^{1/2} \epsilon_{B,-2}^{1/2} \epsilon_{e,-1}^{2},
\end{equation} 
and substituting $\epsilon_e$ from (16) one finds 
$\nu_i \sim {250 \rm eV}\, E_{53}^{-1.07} \epsilon_{B,-5}^{0.36}$. 
These values of $\nu_i$ and $\nu_c$ are consistent with the 
values obtained with detail numerical calculations and reported 
in the Fig. 5 caption.

\begin{figure*}
\begin{centering}
\centerline{\hbox{\includegraphics[width=12cm, angle = 0, clip=true, viewport=.0in .0in 8in 4.5in]{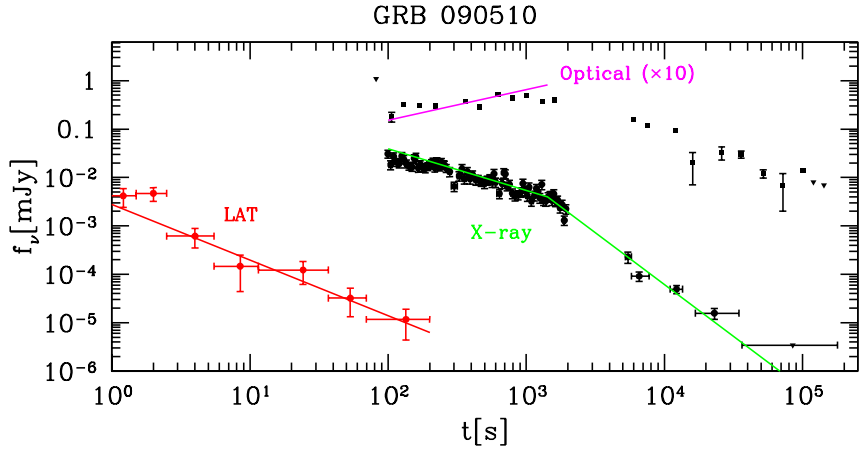}}}
\end{centering}
\caption {Shown in this figure are data for GRB 090510 obtained by {\it Fermi}/LAT
($>$100 MeV), Swift/XRT (X-ray) and Swift/UVOT (optical) data, and a fit to
all these data by the external forward shock model (solid lines). The jet 
break seen in X-ray has been modeled with a power-law, $\propto t^{-p}$; the 
optical light curve after the jet break should show a shallower decay 
$\propto t^{-1/3}$, because at this time $\nu_{opt}<\nu_i$, but then 
it slowly evolves to an asymptotic decay $\propto t^{-p}$ at later times (Rhoads 1999).
The LAT (X-ray) data are from De Pasquale et al. 2009 (Evans et al. 
2007, 2009) and have been converted to flux density at 100 MeV (1 keV) 
using the average spectral index mentioned in the text (\S3.2).  The optical 
data (squares) are from De Pasquale et al. (2010).  Triangles mark upper 
limits in the X-ray and optical light curves.  }
\label{090510-lc}
\end{figure*}

The detailed numerical results of the parameter search can be found in Figure 
\ref{090510-epsilonB}; the sub-space of the 4-D parameter space allowed by
the data for GRB 090510 is projected on the 2-D $\epsilon_B$--$n$ plane,
which is a very convenient way of looking at the allowed sub-space.
Note that all the available data for GRB 090510 can be fitted by the
ES model and that the value of $n$ allowed by the data is less than 
0.1 cm$^{-3}$, which is in keeping with the
low density expected in the neighborhood of short bursts. Moreover,
$\epsilon_B$ for the entire allowed part of the 4-D sub-space is small,
and its magnitude is consistent with what one would expect for the CSM
magnetic field of strength $\lae30\mu$G that is shock compressed by the
blast wave (Fig. \ref{090510-epsilonB}). The ES shock model provides a 
consistent fit to the data from optical to $>$10$^2$MeV bands as 
can be clearly seen in Figure \ref{090510-lc}.  The
ES parameters found for this GRB can be found in the Fig. 5 caption. 
        
\subsection{GRB 080916C}

 The {\it Fermi}/LAT and GBM observations
for this burst have been presented in Abdo et al. (2009a).  For this burst, 
the optical and X-ray observations started about 1d after the burst
and both bands are consistent with 
$f_{\nu}(t)\propto\nu^{-0.5\pm0.3}t^{-1.3\pm0.1}$ (Greiner et al. 2009).

The fact that the optical light curve is decaying as $t^{-1.3}$ means
that $\nu_i$ is below the optical band at 1 day, because if $\nu_i$
is above the optical band, then the light curve should be rising 
as $\propto t^{1/2}$ (as in the case of GRB090510). Moreover, the shallow
spectral index in the Swift/XRT band ($\beta_x < 1$) suggests that $\nu_c>10$keV at 1 day.
The X-ray and optical data together yield a spectral index of $0.65\pm0.03$,
and therefore $p=2.3\pm0.06$ which is consistent with the {\it Fermi}/LAT 
spectrum (see Table 1).
The value of $p$ can be used to calculate the time dependence of the
light curve, and that is found to be $t^{-0.98}$ ($t^{-1.48\pm0.05}$) for 
$s=0$ ($s=2$) CSM. Thus, $s=2$ CSM is preferred 
by the late time optical and X-ray afterglow
data (Kumar \& Barniol Duran 2009; Gao et al. 2009; Zou, Fan \& Piran 2009).  

\begin{figure*}
\begin{centering}
\centerline{\hbox{\includegraphics[width=12cm, angle = 0, clip=true, viewport=.0in .0in 8in 4.5in]{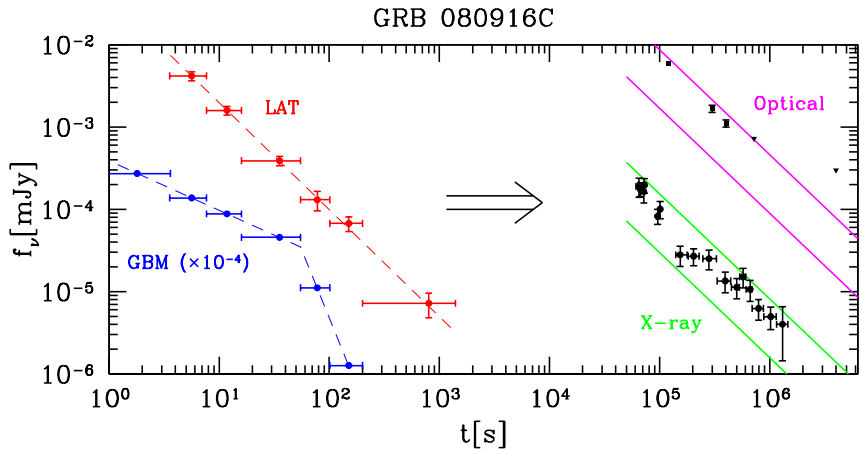}}}
\end{centering}
\caption { The optical and X-ray fluxes of GRB 080916C predicted at late times
using only the high energy data at 150s (assuming synchrotron emission
from external forward shock) are shown in the right half of this figure,
and the predicted flux values are compared with the observed data (discrete
points with error bars).
The width of the region between the green (magenta) lines indicates the 
uncertainty in the theoretically calculated X-ray (optical) fluxes.
The LAT (Abdo et al. 2009a) and X-ray fluxes (Evans et al. 2007, 2009)
at 100 MeV and 1keV, respectively, have been converted to mJy the same way 
as done for Figure \ref{090902B-fwd}.
Optical fluxes (squares) are from Greiner et al. 2009 (triangles are
upper limits). GBM flux at 100keV -- blue filled circles --
is taken from Abdo et al. 2009a.  The thin dashed lines
connecting LAT and GBM data are only to guide the eye.}
\label{080916C-fwd}
\end{figure*}

Using the early $>$100 MeV data only, we determine the ES 
model parameters. With these parameters, we can then predict the X-ray and 
optical fluxes at late times, i.e. the forward direction approach.  The 
constraints that should be satisfied are: (i) The ES flux at 100 MeV and 
150 s should match the observed value (Table 2), (ii) $\nu_c<100$ MeV
to be consistent with the observed spectrum, and (iii) the ES flux at 
150 s should be smaller than the observed value to allow the 100 keV 
flux to decay rapidly as observed.  These constraints are the same 
as the ones presented for the case of GRB090902B and the analytical 
approach is the same as the one presented on \S3.1, therefore, we omit 
the details here.  The ES parameters obtained numerically can be found in 
fig. 2 of Kumar \& Barniol Duran (2009).  With these parameters the X-ray and optical 
flux at late times can be calculated, and we find these in 
excellent agreement with the observations (Figure \ref{080916C-fwd}). 

It is important to note here that this extrapolation from high-energy
early time data to low energy, late time, flux prediction was carried out 
for a circum-stellar medium with $s\sim2$. We have also carried out
the same calculation but for a uniform density medium ($s=0$), and
in this case the theoretically calculated flux at late times is larger than the
observed values by a factor of $\sim$5 or more; the factor of 5 discrepancy is
much larger than error in the flux calculation. We pointed out above that the
late time afterglow data for this burst are consistent with a $s=2$ medium
but not $s=0$ medium. Thus, there is a nice agreement between the late 
time afterglow data and the early $>$10$^2$MeV data --- which explore very
different radii --- in regards to the density stratification of the CSM.

We have carried out the exercise in the ``reverse direction'' as well.  
Using only the late X-ray and optical data, we determine the ES 
parameters. The observational constraints that need to be satisfied 
are: (i) The ES flux at X-ray and optical energies at 1 d should 
match the observed values, (ii) we should have the ordering 
$\nu_i <\nu_{opt}<\nu_X<\nu_c$ to be consistent with the observed spectrum, 
(iii) the ES flux at 150 s should be smaller than the observed value to 
allow the 100 keV flux to decay rapidly as observed, and (iv) the Lorentz Factor 
of the ejecta should be $\gae60$ at 1 d, since we don't want 
$\Gamma$ to be too small at the beginning of the burst, because this would 
contradict estimates done at early times (Greiner et al. 2009).   
Since the analytical approach is very similar to the one for GRB090902B, we omit
it here -- the only difference is that it must be done for a wind-like medium, 
since the data of this GRB prefers it.  The ES parameters can be 
found numerically and with these parameters we predict the $>$100 MeV flux at early 
times.  This predicted flux agrees with the {\it Fermi}/LAT observations as shown in fig. 
3 of Kumar \& Barniol Duran (2009).

\section{Discussion and Conclusion}    

The {\it Fermi} Satellite has detected 12 GRBs with $>$100 MeV emission in 
about one year of operation.  In this paper we have analyzed 
the $>$100 MeV emission of three of them: two long-GRBs
(090902B and 080916C) and one short burst (GRB 090510), and find that the
data for all three bursts are consistent with synchrotron emission in the
external forward shock.  This idea was initially proposed in our
previous work on GRB 080916C (Kumar \& Barniol Duran 2009), shortly
after the publication of this burst's data by Abdo et al. (2009a).
Now, there are three GRBs for which high energy data has been published, and
for all of them we have presented here multiple lines of evidence
that $>$100 MeV photons, subsequent to the prompt GRB phase, were generated
in the external forward shock. The reason that high energy photons are 
detected from only a small fraction of GRBs observed by {\it Fermi} is likely 
due to the fact that the high energy flux from the external forward 
shock has a strong dependence on the GRB jet Lorentz factor, and therefore
very bright bursts with large Lorentz factors are the only ones detected 
by {\it Fermi}/LAT (this was pointed out by Kumar \& Barniol Duran 2009, who 
also suggested that there should be no difference in long and short 
bursts, as far as the $>$100 MeV emission is concerned - the high 
energy flux is only a function of burst energy and time, eq. 1).

We have analyzed the data in 4 different ways, and all of them lead to the
same conclusion regarding the origin of $>$10$^2$MeV photons. First,
we verified that the temporal decay index for the $>$100 MeV
light curve and the spectral index are consistent
with the closure relation expected for the synchrotron
emission in the external forward shock. Second, we calculated
the expected magnitude of the synchrotron flux at 100 MeV according to
the external forward shock model and find that to be consistent with
the observed value.  Third, using the $>$100 MeV data only, we determined the
external shock parameters, and from these parameters we predict the X-ray
and optical fluxes at late times and find that these predicted fluxes
are consistent with the observed values within the uncertainty of our
calculations, i.e. a factor of two (see figs. \ref{090902B-fwd},
\ref{080916C-fwd}). And lastly, using the late time X-ray,
optical and radio fluxes --- which the GRB community has believed for a long
time to be produced in the external forward shock --- we determine the external
shock parameters, and using these parameters we predict the expected
$>$100 MeV flux at early times and find the flux to be in agreement
with the observed value (see fig. \ref{090902B-reverse}). The 
fact that the $>$100 MeV emission and the lower energy ($\lae 1$ MeV)
emission are produced by two different sources at two different locations
suggests that we should be cautious when using the highest observed photon energy and 
pair-production arguments to determine the Lorentz factor of the GRB jet.

We point out that the external shocks for these bursts were nearly adiabatic,
i.e. radiative losses are small. The evidence for this comes from two different
observations: (1) the late time X-ray spectrum lies in the adiabatic regime;
(2) a radiative shock at early times (close to the deceleration time) would
produce emission in the 10--10$^2$ keV band far in excess of the observed
limits. We find that radiative shock is not needed to explain the 
temporal decay index of the $>$100 MeV light curve as suggested by 
(Ghisellini, Ghirlanda, Nava 2010), provided that the 
observing band is above all synchrotron characteristic frequencies. 

We find that the magnetic field required in the external forward shock
for the observed high and low energy emissions for these three bursts is
consistent with shock-compressed magnetic field in the CSM; 
the magnetic field in the CSM -- before shock compression -- should
be on the order of a few tens of micro-Gauss (see figs.
\ref{090902B-epsilonB}, \ref{090902B-epsilonBlate} and \ref{090510-epsilonB}).
For these three bursts, at least, no magnetic dynamo is needed to operate
behind the shock front to amplify the magnetic field.

The data for the short burst (GRB 090510) are consistent with the medium in
the vicinity of the burst (within $\sim$1 pc) being uniform and with
density less than 0.1 cm$^{-3}$; the data rules out a CSM where 
$n\propto R^{-2}$. On the other hand, the data for one of the
two long {\it Fermi} bursts (GRB 080916C) prefers a wind like medium and
the other (GRB 090902B) a uniform density medium; these conclusions are
reached independently from late time afterglow data alone and from
the early time high energy data projected to late time using the
4-D parameter space technique described in \S3.

It is also interesting to note that the power-law index of the 
energy distribution of injected electrons ($p$) in the shocked fluid, for all the
three {\it Fermi} bursts analyzed in this work, is $2.4$ to within the error of
measurement, suggesting an agreement with the {\it Fermi} acceleration of particles
in highly relativistic shocks, e.g. Achterberg et al. (2001);  
a unique power-law index for electrons' distribution in highly relativistic shocks 
is not found in all simulations.  The study of high energy emission close to the deceleration 
time of GRB jets is likely to shed light on the onset of collisionless shocks and 
particle acceleration process.

It might seem surprising that we are able to fit all data (optical, X-ray,
$\lae10^2$MeV) for these three {\it Fermi} bursts with just a few parameters for
the external forward shock. This is in sharp contrast to Swift bursts
which often display a variety of puzzling (poorly understood) features in
their afterglow light curves. There are two reasons that these {\it Fermi} bursts
can be understood using a very simple model (external forward shock). 
(1) The data for the two long {\it Fermi} bursts (080916C and 090902B) are not available during the
first 1/2 day, and that is precisely the time frame when complicated
features (plateau, etc., eg. Nousek et al. 2006, O'Brien et al. 2006) are 
seen in the X-ray afterglow light curves of
Swift bursts (we note that the external forward shock model 
in its simplest form can't explain these features) --- however,
the afterglow data at later times is almost invariably
a smooth single (or double) power-law function that can be modeled by
synchrotron emission from an external forward shock. (2) For very
energetic GRBs --- the three bursts we have analyzed in this paper are
among the brightest bursts in their class --- the progenitor star is
likely to be completely destroyed leaving behind very little
material to fall back onto the compact remnant at the center
to fuel continued activity and give rise to complex
 features during the first few hours of the X-ray afterglow light curve 
(Kumar, Narayan \& Johnson, 2008).
To summarize, the GRB afterglow physics was simple in the decade
preceding the launch of Swift, and then things became quite complicated,
and now the {\it Fermi} data might be helping to clear the fog and
reveal the underlying simplicity once again.

\section*{Acknowledgments}

RBD thanks Massimiliano De Pasquale for clarifying 
some aspects of the data of GRB 090510.  We also 
thank Alex Kann for useful discussions about the optical
data.  We thank the referee for a constructive report.
This work has been funded in part by NSF grant ast-0909110.
This work made use of data supplied by the UK Swift
Science Data Centre at the University of Leicester.

\end{document}